\documentclass[preprint,12pt]{elsarticle}

\usepackage{amssymb}

\usepackage{graphicx,color}
\usepackage[usenames,dvipsnames]{xcolor}
\usepackage[section]{placeins}
\usepackage[normalem]{ulem}

\usepackage{natbib}

\journal{Journal of Subatomic Particles and Cosmology}

\usepackage{amssymb}
\usepackage{amsmath}
\usepackage{ytableau}

\usepackage{multirow}
\usepackage{epsfig,color,graphicx}
\usepackage{xcolor}
\usepackage{subfigure}

\colorlet{darkred}{red!80!black}
\colorlet{darkgreen}{green!50!black}
\colorlet{darkblue}{blue!50!black}

\usepackage{soul}

\usepackage[pdftex,hypertexnames=false,linktocpage=true]{hyperref}
\hypersetup{colorlinks=true,linkcolor=blue,anchorcolor=blue,citecolor=blue,filecolor=blue,urlcolor=black,bookmarksnumbered=true,
pdfview=FitB
}

\begin{document}

\begin{frontmatter}
\title{Basis light-front quantization approach to deuteron}

\author[imp,ucas]{Chandan Mondal\corref{speaker}}
\ead{mondal@impcas.ac.cn}

\author[imp,ucas]{Satvir Kaur}
\ead{satvir@impcas.ac.cn}

\author[imp,ucas]{Jiatong Wu}
\ead{wujt@impcas.ac.cn}

\author[imp,ucas,iowa]{Siqi Xu}
\ead{xsq234@impcas.ac.cn}

\author[imp,ucas]{Xingbo Zhao}
\ead{xbzhao@impcas.ac.cn}

\author[iowa]{James~P.~Vary}
\ead{jvary@iastate.edu}

\author[]{\\\vspace{0.2cm}(BLFQ Collaboration)}

\address[imp]{Institute of Modern Physics, Chinese Academy of Sciences, Lanzhou, Gansu, 730000, China}
\address[ucas]{School of Nuclear Physics, University of Chinese Academy of Sciences, Beijing, 100049, China}
\address[affsch]{Affiliated School of Huizhou University, Huizhou, Guangdong, 516001, China}
\address[iowa]{Department of Physics and Astronomy, Iowa State University, Ames, IA 50011, USA}
\cortext[speaker]{Speaker.}

\begin{abstract}
We obtain the deuteron’s wave functions as eigenstates of the light-front quantum chromodynamics (QCD) Hamiltonian  using a fully relativistic and nonperturbative approach based on light-front quantization, without an explicit confining potential. These eigenstates include six-quark and six-quark–one-gluon components. The deuteron wave function consists of both a singlet-singlet color state and additional hidden color states arising from non-trivial color rearrangements. Our results reveal that while the singlet-singlet state is present, the hidden color states collectively dominate, contributing a larger probability to the deuteron wave function. This highlights the significant role of hidden color components in the QCD description of nuclear structure. Using these wave functions, we investigate the deuteron’s electromagnetic properties.

\end{abstract}
\end{frontmatter}

\section{Introduction}
The deuteron, as the simplest nuclear bound state, serves as an ideal testbed for exploring the interplay between quantum chromodynamics (QCD) and nuclear structure. While it is traditionally modeled as a loosely bound proton-neutron system dominated by meson exchange, a more fundamental QCD perspective describes it primarily as a six-quark system, where color degrees of freedom play a crucial role in shaping its internal structure.

In QCD, quarks interact via the strong force, mediated by gluons and governed by SU(3) color symmetry. Each quark carries a color charge—red, green, or blue—while gluons facilitate interactions by exchanging color among them. Due to color confinement, quarks and gluons must combine into overall color-neutral states. Unlike baryons, which form simple color-singlet configurations, the deuteron, as a six-quark system, allows for more intricate color structures, known as hidden-color states. The significance of these hidden-color contributions has been widely discussed in the literature~\cite{Matveev:1977xt, Matveev:1977ha, Hogaasen:1979qa, Brodsky:1983vf, Farrar:1991qi, Bashkanov:2013cla, Miller:2013hla, Bakker:2014cua}. Though traditionally overlooked in nuclear models, these states can play a crucial role in the deuteron wave function, particularly in short-range nuclear correlations and deep inelastic scattering observables.

A nucleon-based description of the deuteron struggles to capture its intricate color structures. At high momentum transfers or short-distance scales, where quark and gluon dynamics become significant, traditional nucleon-nucleon models lose validity. Examining the deuteron at the quark level is therefore essential for a deeper understanding of its internal structure and its role in high-energy nuclear interactions.

Unlike spin-$0$ and spin-$1/2$ systems, the deuteron’s spin configuration features distinctive properties, including the tensor force, magnetic dipole moment, and electric quadrupole moment. These attributes make it a crucial system for investigating fundamental nuclear forces and spin-dependent aspects of QCD.  

The electromagnetic properties of the deuteron have been extensively studied using various theoretical approaches~\cite{Gutsche:2016lrz, Gutsche:2015xva, Sun:2016ncc, Shi:2022blm, Huseynova:2022tok, Marcucci:2015rca, Khachi:2023iqp, Chemtob:1974nf, Blankenbecler:1971xa, Piller:1995mf}, and its form factors have been extracted from global analyses of elastic electron-deuteron scattering data~\cite{JLABt20:2000qyq, Sick:1998cvq, Zhang:2011zu, Nikolenko:2003zq}. For a comprehensive review of these experimental results, see Refs.~\cite{Marcucci:2015rca, Garcon:2001sz}. However, relatively little research has explored the quantum spin structure of the deuteron at the level of its fundamental constituents. While the overall spin of the deuteron is well established, the internal mechanisms contributing to and sustaining this spin—such as quark orbital motion, gluonic effects, and sea-quark contributions—remain incompletely understood.  

Recently, the unpolarized parton distribution function of a deuteron-like dibaryon system was investigated for the first time using lattice QCD~\cite{Chen:2024rgi}. As a spin-1 nucleus, the deuteron also exhibits the tensor-polarized structure function $b_1$, in addition to the conventional structure functions observed in spin-1/2 systems. Since its initial measurement by HERMES in 2005~\cite{HERMES:2005pon}, no further experiments have been conducted, but proposals from Jefferson Lab~\cite{ProposalJLab2023} and capabilities at Fermilab~\cite{AbilityFermiLabE1039, Keller:2022abm} offer promising opportunities to explore this unique observable.  

Theoretical studies suggest that $b_1$ arises from an intricate interplay between nucleonic and nuclear effects. Unlike standard structure functions, which reflect intrinsic quark distributions within nucleons, $b_1$ is also influenced by the collective spin configuration of the two nucleons in the deuteron, highlighting a distinctly nuclear property. This connection between nucleon and nuclear structure provides a valuable avenue for probing the complexity of the deuteron~\cite{AbilityFermiLabE1039}.  

Interestingly, predictions for $b_1$ show discrepancies with HERMES data, signaling gaps in our understanding of the deuteron's internal dynamics~\cite{Hoodbhoy:1988am, Khan:1991qk, Edelmann:1997ik, Cosyn:2017fbo}. Addressing these inconsistencies could pave the way for new discoveries in spin physics, extending beyond our current theoretical framework~\cite{Kumano:2024fpr}. While data from HERMES remains limited, future experiments at next-generation facilities such as the Electron-Ion Collider (EIC) in the USA and the proposed EicC in China~\cite{Accardi:2012qut, Anderle:2021wcy} are expected to provide crucial insights into the partonic structure of light nuclei.

The basis light-front quantization (BLFQ) approach~\cite{Vary:2009gt} has been successfully applied to various QCD systems, from mesons to baryons, offering a non-perturbative framework to solve QCD from first principles. By expanding the Fock space boundaries, BLFQ has provided valuable insights into hadronic properties~\cite{Lan:2019vui, Lan:2019rba, Mondal:2019jdg, Lan:2019img, Lan:2021wok, Xu:2021wwj, Liu:2022fvl, Hu:2022ctr, Peng:2022lte, Xu:2022yxb, Zhu:2023lst, Zhu:2023nhl, Kaur:2023lun, Lin:2023ezw, Zhang:2023xfe, Kaur:2024iwn, Liu:2024umn, Yu:2024mxo, Nair:2024fit, Zhu:2024awq, Wu:2024hre, Lin:2024ijo, Xu:2024sjt, Peng:2024qpw, Lan:2025fia, Zhang:2025nll}. Building on this success, we extend BLFQ to study the deuteron—the lightest nuclear bound state—enabling a direct investigation of its wave function in terms of quark and gluon degrees of freedom. This approach offers a novel perspective on the deuteron’s color structure beyond the conventional nucleon-nucleon framework.

In this work, we apply the BLFQ framework to investigate the probability distribution of different color configurations within the deuteron wave function. As an initial study, we truncate the Fock space to include only the six-quark (\(qqqqqq\)) and six-quark–one-gluon (\(qqqqqq~g\)) components. By solving the light-front QCD (LFQCD) Hamiltonian in this truncated basis for its mass eigenstates and fitting the quark masses and coupling constant, we obtain the deuteron light-front wave functions (LFWFs). We then use these LFWFs to compute the probabilities of various color states, distinguishing between the conventional nucleon-nucleon (singlet-singlet) configuration and hidden-color states arising from non-trivial SU(3) color rearrangements. Our results demonstrate that hidden-color states play a substantial role in the deuteron wave function. Finally, we employ the deuteron LFWFs to analyze its electromagnetic properties.

%In this work, we apply the BLFQ framework to investigate the probability distribution of different color configurations within the deuteron wave function. As an initial study, we truncate the Fock space to include only the six-quark ($qqqqqq$) and six-quark–one-gluon ($qqqqqq~g$) components. By solving the light-front QCD Hamiltonian in this truncated basis for its mass eigenstates and and fitting the quark masses and coupling constant, we obtain the deuteron light-front wave functions (LFWFs). We then employ the LFWFs to compute the probabilities of various color states, distinguishing between the conventional nucleon-nucleon (singlet-singlet) configuration and hidden-color states arising from non-trivial SU(3) color rearrangements. Our results demonstrate that hidden-color states play a substantial role in the deuteron wave function. Using the resulting deuteron LFWFs, we further analyze its electromagnetic properties.

\section{Deuteron wave functions from light-front QCD Hamiltonian}

In the BLFQ framework, we solve the eigenvalue problem of the LFQCD Hamiltonian:
\begin{equation}  
P^+ P^- |\Psi\rangle = M^2 |\Psi\rangle,  
\end{equation}  
where \( P^+ \) represents the longitudinal momentum, and \( P^- \) is the light-front Hamiltonian acting on the deuteron state \( |\Psi\rangle \). The deuteron wave function is expanded in Fock space, incorporating various partonic components consisting of quarks (\( q \)), antiquarks (\( \bar{q} \)), and gluons (\( g \)) \cite{Brodsky:1997de}:  
\begin{equation}  \label{Fock}
|\Psi\rangle = \psi_{(qqq\,qqq)} |qqq~qqq\rangle + \psi_{(qqq\,qqq\,g)} |qqq\,qqq\,g\rangle + \psi_{(qqq\,qqq\,q\bar{q})} |qqq\,qqq\,q\bar{q}\rangle + \dots  
\end{equation}  
where \( \psi(...) \) represents the probability amplitudes of different Fock-state components. In this work, we restrict our analysis to the first two terms in Eq.~\eqref{Fock}, namely the six-quark (\( |qqq\,qqq\rangle \)) and six-quark–one-gluon (\( |qqq\,qqq\,g\rangle \)) configurations.  

The LFQCD Hamiltonian with a dynamical gluon in the light-front gauge (\( A^+ = 0 \)) is given by \cite{Brodsky:1997de}:  
\begin{align}  
P^-_{\text{QCD}} = \int d^2x_{\perp} dx^- &\Bigg[  
\frac{1}{2} \bar{\psi} \gamma^+ \frac{m_0^2 + (i\partial^{\perp})^2}{i\partial^+} \psi  
+ A^\mu_a \frac{m_g^2 + (i\partial^{\perp})^2}{2} A_\mu^a  
+ g_s \bar{\psi} \gamma^\mu T^a A_\mu^a \psi  \nonumber\\  
&  
+ \frac{g_s^2}{2} \bar{\psi} \gamma^+ T^a \psi \frac{1}{(i\partial^+)^2} \bar{\psi} \gamma^+ T^a \psi  
\Bigg].  
\end{align}  
The first two terms describe the kinetic energy contributions, where \( m_0 \) and \( m_g \) are the bare masses of the quark and the gluon, respectively. The fields \( \psi \) and \( A^\mu \) represent the quark and gluon fields, while \( x^- \) and \( x_{\perp} \) denote the longitudinal and transverse coordinates. The last two terms correspond to the quark-gluon interactions and four-fermion interactions, with \( g_s \) as the strong coupling constant. The SU(3) generators \( T^a \) and the Dirac matrices \( \gamma^\mu \) encode the color and spin structures of QCD interactions.
 
To incorporate quark mass corrections due to quantum fluctuations to higher Fock sectors, we introduce a mass counter term, $\delta m_q = m_0 - m_q$, for the quark in the leading Fock component, where $m_q$ is the renormalized quark mass \cite{Glazek:1993rc, Karmanov:2008br,Karmanov:2012aj, Zhao:2014hpa, Zhao:2020kuf}. Additionally, a mass parameter $m_f$ is introduced to parameterize nonperturbative effects in the vertex interactions \cite{Glazek:1992aq, Burkardt:1998dd}. In our current Fock sector truncation, there are no mass corrections for the gluon and the physical gluon mass is $m_g=0$.

In the BLFQ framework, the longitudinal and transverse dynamics of Fock particles are described using a discretized plane-wave basis and two-dimensional (2D) harmonic oscillator (HO) wave functions, respectively. The longitudinal motion is confined to a one-dimensional box of length $2L$, with antiperiodic (periodic) boundary conditions imposed on fermions (bosons). The longitudinal momentum is discretized as $p^+ = \frac{2\pi k}{L}$, where $k$ takes half-integer values for fermions and integer values for bosons, with the bosonic zero mode explicitly excluded. The total longitudinal momentum is given by $P^+ = \frac{2\pi K}{L}$, where $K = \sum_i k_i$, and the longitudinal momentum fraction of the $i$-th parton is defined as $x_i = p_i^+/P^+ = k_i/K$.

For the transverse degrees of freedom, we employ 2D HO wave functions $\phi_{n,m}(k_\perp; b)$, characterized by radial ($n$) and angular momentum ($m$) quantum numbers, with $b$ representing the HO scale parameter. A single-particle basis state is fully specified by the set of quantum numbers $\alpha = \{x, n, m, \lambda\}$, where $\lambda$ denotes helicity. Many-body basis states are constructed as direct products of single-particle states, $|\alpha \rangle = \otimes_i |\alpha_i \rangle$, with the total angular momentum projection given by $m_J = \sum_i (m_i + \lambda_i)$. The deuteron Fock sector decomposition includes multiple color-singlet configurations: five in the first Fock sector and sixteen in the second.

The basis truncation is governed by two parameters: $K$ for longitudinal resolution and $N_{\text{max}}$ for transverse excitations. The transverse truncation condition $N_{\text{max}} \geq 2n_i + |m_i| + 1$ introduces effective ultraviolet ($\Lambda_{\text{UV}} = b\sqrt{N_{\text{max}}}$) and infrared ($\Lambda_{\text{IR}} = b/\sqrt{N_{\text{max}}}$) cutoffs, regulating high- and low-momentum regimes, respectively.

The deuteron LFWFs with helicity $\Lambda$ in momentum space are expressed as superpositions of Fock sector components:
\begin{equation}
\Psi^{\mathcal{N},\,\Lambda}_{\{x_i,\vec{p}_{\perp i},\lambda_i\}} 
= \sum_{\{n_i, m_i\}} \psi^{\mathcal{N}}(\{\overline{\alpha}_i\}) 
\prod_{i=1}^{\mathcal{N}} \phi_{n_i m_i}(\vec{p}_{\perp i},b),
\label{eqn:wf}
\end{equation}
where $\psi^{\mathcal{N}}(\{\overline{\alpha}_i\})$ represents the amplitude for the $\mathcal{N}$-particle Fock sector, with $\mathcal{N}$ corresponding to either the $|qqq\,qqq\rangle$ or $|qqq\,qqq\,g\rangle$ configuration. These wave function components are obtained by diagonalizing the full Hamiltonian matrix in the BLFQ framework.

For our current numerical calculations, we use $N_{\text{max}} = 8$ and $K = 9$, with the harmonic oscillator scale parameter set to $b = 0.30$ GeV. We fix the instantaneous interaction cutoff at $b_{\text{inst}} = 5.00$ GeV. The model parameters $\{m_u, m_d, m_f, g_s\} = \{1.00, 0.95, 42.56, 1.90\}$ (with all masses in GeV except the dimensionless coupling constant $g_s$) were determined by fitting to both the deuteron mass and its electromagnetic properties. The relatively large constituent quark masses in the dominant Fock sectors serve two purposes: they partially account for confinement effects, and make substantial contributions to the deuteron mass as a QCD bound state.

\section{Color structure}
In this work, we model the deuteron within a Fock space framework, where the first Fock sector consists of six valence quarks and the second Fock sector contains six quarks plus one dynamical gluon. Within SU(3) color symmetry, the six-quark Fock state decomposes according to the group representation~\cite{Bakker:2014cua}:
\begin{align}
{\tiny \ydiagram{1} \otimes \ydiagram{1} \otimes \ydiagram{1} \otimes \ydiagram{1} \otimes \ydiagram{1} \otimes \ydiagram{1}}&= 
{\tiny \ydiagram{6} \oplus 5 ~~\ydiagram{5,1} \oplus 9 ~~\ydiagram{4,2} \oplus 10 ~~\ydiagram{4,1,1}} \nonumber\\
& {\tiny \oplus ~ 5 ~~\ydiagram{3,3} \oplus 16~~\ydiagram{3,2,1} \oplus 5~~\ydiagram{2,2,2}}
\end{align} 
\begin{align}
    \{ 3 \} \otimes \{ 3 \} \otimes \{ 3 \} \otimes \{ 3 \} \otimes \{ 3 \} \otimes \{ 3 \} &= 729 \nonumber\\ 
     &=1~\{28\}\oplus 5~ \{35 \} \oplus 9~ \{ 27 \} \oplus 10 ~\{10\} \nonumber\\
    &\oplus 5~ \{ 10 \} \oplus 16 ~\{8 \} \oplus {\color{blue}5~ \{1 \}}
\end{align}
Among the five color-singlet states (highlighted in blue) in the six-quark configuration, one state is a pure singlet arising from the singlet-singlet configuration, while the remaining four represent hidden color states derived from the octet-octet representation.

The allowed color configurations for the deuteron's second Fock sector (comprising six quarks and one dynamical gluon) can be expressed as:
\begin{align}
{\tiny \ydiagram{1} \otimes \ydiagram{1} \otimes \ydiagram{1} \otimes \ydiagram{1} \otimes \ydiagram{1} \otimes \ydiagram{1} \otimes \ydiagram{2,1}} 
&= {\tiny \ydiagram{8,1} \oplus 6~~\ydiagram{7,2} \oplus 6~~\ydiagram{7,1,1}} \nonumber\\
&{\oplus \tiny ~ 14~~\ydiagram{6,3} \oplus  30~~\ydiagram{6,2,1} \oplus 14~~\ydiagram{5,4}} \nonumber\\
& {\tiny \oplus ~54~~\ydiagram{5,3,1} \oplus 40~~\ydiagram{5,2,2} \oplus 30~~\ydiagram{4,4,1}} \nonumber\\
& {\tiny\oplus~ 61~~\ydiagram{4,3,2}  \oplus~ 16 ~~\ydiagram{3,3,3}}
\end{align}
 \begin{align}
     \{ 3 \} \otimes \{ 3 \} \otimes \{ 3 \} \otimes \{ 3 \} \otimes \{ 3 \} \otimes \{ 3 \} \otimes \{8 \} =&~ 5832 \nonumber\\
     =& ~ \{80 \} \oplus 6 ~\{81 \} \oplus 6~ \{28\} \oplus 14~\{64\} \oplus 30~\{35\} \nonumber\\
     \oplus&~ 14~\{35\} \oplus 54~\{27\} \oplus 40~\{10 \} \oplus 30~\{10\} \nonumber\\
     \oplus&~ 61~\{8\} \oplus {\color{blue}16~\{1\}}
 \end{align}
In the second Fock sector of the deuteron, we identify 16 hidden color-singlet states (highlighted in blue). Consequently, the deuteron in our framework comprises a total of $21$ color-singlet states: one pure state originating from the singlet-singlet combination of the two nucleons, and the remaining $20$ classified as hidden color states. The specific color configurations of the $|qqq\,qqq\rangle$ and $|qqq\,qqq\,g\rangle$ Fock components that give rise to these color-singlet states are detailed in Table~\ref{Table:color}.

To incorporate essential QCD features into the deuteron calculations, we compute the color factors by evaluating the matrix elements of $T^a$ and $T^a T^a$ between the color-singlet wave functions. These factors serve as multiplicative terms in the matrix elements of the LFQCD Hamiltonian within the BLFQ framework.

% \begin{table}[hbt!]
% \caption{The color singlet states corresponding to the different configurations in SU(3) color symmetry representation.}
% \centering
% \begin{tabular}{|l|l|c|c|c|}
%   \hline
%   & {\bf Color Configurations} & {\bf Total Color Singlet States} & \multicolumn{2}{c|}{\bf Probability (\%)} \\
%   & & & \multicolumn{2}{c|}{(Preliminary)} \\
%   \cline{4-5}
%   & & & {\bf \(m_J = 0\)} & {\bf \(m_J = 1\)} \\ \hline
%   \multirow{$| qqq~qqq \rangle$}  
%   & Singlet-Singlet & 1 & 44.56 & 44.52 \\  
%   & Octet-Octet & 4 & 12.98 & 13.02 \\ \hline
%   & Decuplet-Octet-Octet & 2 &  &  \\  
%   & Octet-Decuplet-Octet & 2 &  &  \\  
%   \multirow{$| qqq~qqq~g \rangle$} & Octet-Octet-Octet & 8 & 42.46 &  42.46\\  
%   & Octet-Singlet-Octet & 2 &  &  \\  
%   & Singlet-Octet-Octet & 2 &  &  \\ \hline
% \end{tabular}
% \label{Table:color}
% \end{table}

\begin{table}[hbt!]
\caption{The color singlet states corresponding to the different configurations in SU(3) color symmetry representation. The probabilities are obtained in our BLFQ approach with $N_{\text{max}} = 8$ and $K = 9$.}
\vspace*{0.15cm}
\centering
\begin{tabular}{|l|l|c|c|c|}
  \hline
  & {\bf Color Configurations} & {\bf Total Color Singlet States} & \multicolumn{2}{c|}{\bf Probability (\%)} \\
  & & & \multicolumn{2}{c|}{(Preliminary)} \\
  \cline{4-5}
  & & & {\bf \(m_J = 0\)} & {\bf \(m_J = 1\)} \\ \hline
  \multirow{2}{*}{$| qqq~qqq \rangle$}  
  & Singlet-Singlet & 1 & 44.56 & 44.52 \\  
  & Octet-Octet & 4 & 12.98 & 13.02 \\ \hline
  \multirow{2}{*}{ } & Decuplet-Octet-Octet & 2 &  &  \\  
  & Octet-Decuplet-Octet & 2 &  &  \\  
  \multirow{3}{*}{$| qqq~qqq~g \rangle$} & Octet-Octet-Octet & 8 & 42.46 & 42.46 \\  
  & Octet-Singlet-Octet & 2 &  &  \\  
  & Singlet-Octet-Octet & 2 &  &  \\ \hline
\end{tabular}
\label{Table:color}
\end{table}

\begin{figure}[hbt!]
    \centering
    \includegraphics[width=0.8\linewidth]{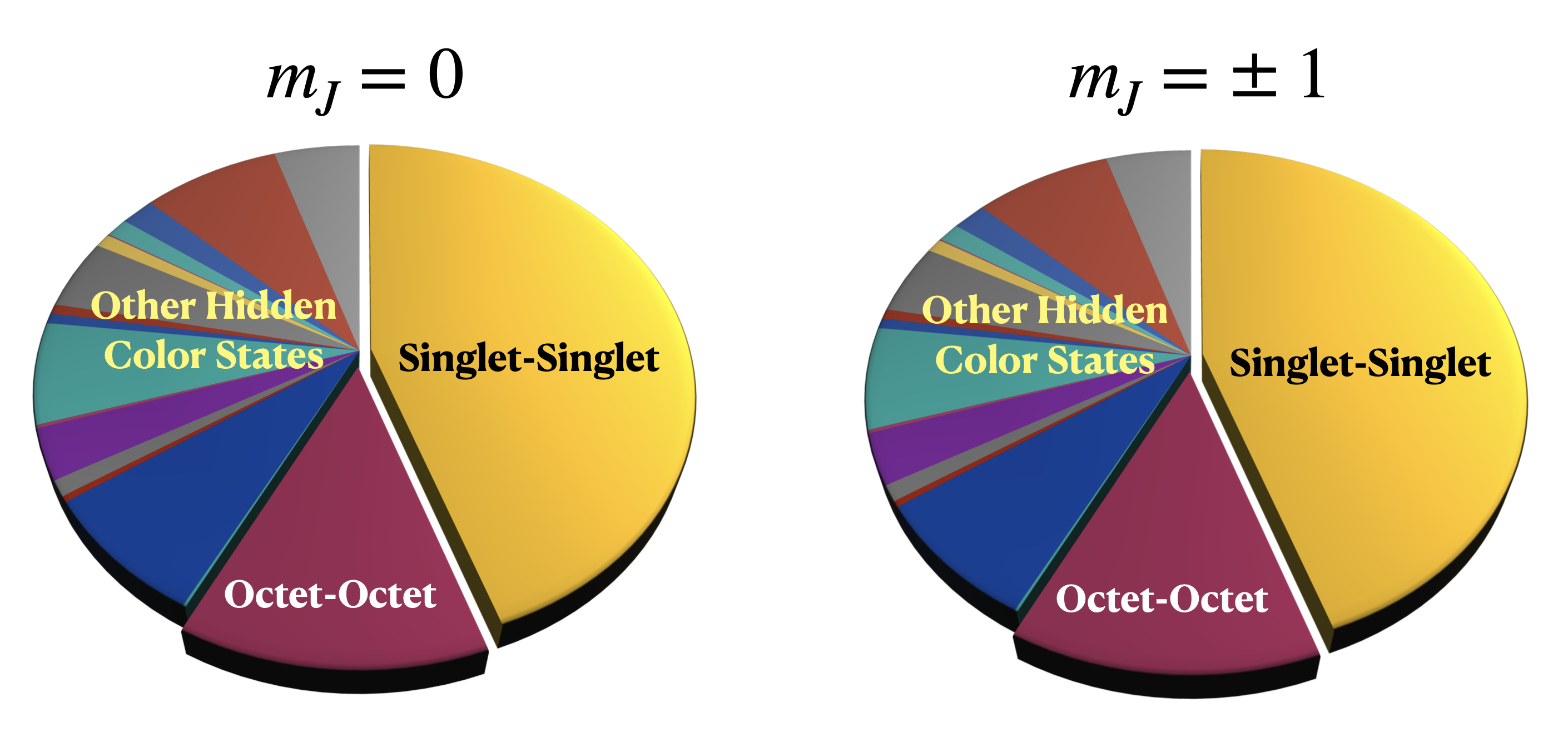}
    \caption{Probability of different color configurations within deuteron for $m_J=0$ and $m_J=\pm 1$ spin projections in our BLFQ approach with $N_{\text{max}} = 8$ and $K = 9$.}
    \label{fig:probability}
\end{figure}
Using the LFWFs obtained from Hamiltonian diagonalization, we evaluate the probability distribution across all color states to quantify their individual contributions. Our results demonstrate that while the pure color state makes a substantial contribution (44.5\%) at our model scale, the hidden color states collectively dominate with 55.5\% of the total probability. This distribution remains consistent for both $m_J = 0$ and $m_J = \pm 1$ components in current scenario, with only negligible variations observed between different spin projections. The complete probability distribution is presented in Table~\ref{Table:color}, and Fig.~\ref{fig:probability} provides a visual comparison of these contributions across different spin configurations.

\section{Electromagnetic form factors}
The electromagnetic form factors (EMFFs) of a spin-$1$ target are defined through the hadronic matrix elements of the electromagnetic current $J^{\mu}$ between the target's initial and final states~\cite{Arnold:1979cg}.
\begin{align}
\langle P^\prime,\Lambda^\prime|J^{\mu}|P,\Lambda\rangle = &-\epsilon^{\ast}_{\Lambda^{\prime}}\cdot\epsilon_{\Lambda}(P+P^{\prime})^{\mu}F_{1}(Q^{2}) +\left(\epsilon^{\mu}_{\Lambda}q\cdot\epsilon^{\ast}_{\Lambda^{\prime}}-\epsilon^{\ast\mu}_{\Lambda^{\prime}}q\cdot\epsilon_{\Lambda}\right)F_{2}(Q^{2}) \nonumber \\
&+\frac{(\epsilon^{\ast}_{\Lambda^{\prime}}\cdot q)(\epsilon_{\Lambda}\cdot q)}{2M_{D}^{2}}(P+P^{\prime})^{\mu}F_{3}(Q^{2}), 
\end{align}
where $q = P^{\prime} - P$ and $Q^2 = -q^2 = \vec{q}_{\perp}^{\,2}$; $\epsilon_{\Lambda}$ and $\epsilon_{\Lambda^{\prime}}$ represent polarization vectors; and $F_{1,2,3}$ are Lorentz-invariant form factors dependent on the target mass $M_{D}$. 

We evaluate these matrix elements in the Breit frame with specific kinematics~\cite{Cardarelli:1994yq,Brodsky:1992px}:
\begin{equation}
\begin{aligned}
q^{\mu}&=(0,0,Q,0), \\
P^{\mu}&=(M_{V}\sqrt{1+\eta},M_{D}\sqrt{1+\eta},-Q/2,0), \\
P^{\prime\mu}&=(M_{V}\sqrt{1+\eta},M_{D}\sqrt{1+\eta},Q/2,0),
\end{aligned}
\end{equation}
where $\eta=Q^{2}/4M_{D}^{2}$.
On the light front, we compute the form factors using the plus component of the quark electromagnetic current, $J^+_q(0)$, which is given in terms of its matrix elements as~\cite{Qian:2020utg,Li:2021cwv}.
\begin{align}
I_{\Lambda^\prime, \Lambda}^{+} (Q^2) &\equiv \left\langle P^\prime,\Lambda^\prime\left| \frac{J^+_q(0)}{2P^+}\right| P,\Lambda \right\rangle \nonumber \\
&= \int_{\mathcal{N}} \Psi^{\mathcal{N},\,\Lambda^\prime\,*}_{\{x_i^\prime,\vec{p}_{\perp i}^{\,\prime},\lambda_i\}} \, \Psi^{\mathcal{N},\,\Lambda}_{\{x_i,\vec{p}_{\perp i},\lambda_i\}}, \label{eq:FFs}
\end{align}
where the integration measure $\int_{\mathcal{N}}$ is defined as:
\[
\int_{\mathcal{N}} \equiv \sum_{\mathcal{N},\,\lambda_i} \prod_{i=1}^\mathcal{N} \int \left[\frac{{\rm d}x\,{\rm d}^2\vec{p}_\perp}{16\pi^3}\right]_i 16\pi^3 \, \delta\left(1-\sum x_j\right) \, \delta^2\left(\sum \vec{p}_{\perp j}\right).
\]
The struck quark's momentum transforms as $x'_1 = x_1$ and $\vec{p}_{\perp 1}' = \vec{p}_{\perp 1} + (1-x_1)\vec{q}_\perp$, while spectator partons follow $x'_i = x_i$ and $\vec{p}_{\perp i}' = \vec{p}_{\perp i} - x_i\vec{q}_\perp$.
Note that nine possible helicity combinations $\Lambda, \Lambda^{\prime} = 0, \pm 1$ reduce to four independent elements ($I_{1,1}^{+},\,I_{1,-1}^{+},\,I_{1,0}^{+},\,I_{0,0}^{+}$) through light-front symmetry constraints~\cite{Cardarelli:1994yq}. 
%By considering all possible combinations of incoming and outgoing vector meson helicities, $\Lambda, \Lambda^{\prime} = 0, \pm 1$, one obtains nine matrix elements of the electromagnetic current, $I^{+}_{\Lambda^{\prime},\Lambda}$. However, using light-front parity and time-reversal invariance, these nine elements can be reduced to four independent ones: $I_{1,1}^{+},\,I_{1,-1}^{+},\,I_{1,0}^{+}$, and $I_{0,0}^{+}$~\cite{Cardarelli:1994yq}.

The physical form factors --- charge ($G_C$), magnetic ($G_M$), and quadrupole ($G_Q$) ---relate to the Lorentz-invariant ones as~\cite{Choi:2004ww,Brodsky:1992px}:
\begin{align}
G_{C}=F_{1}+\frac{2}{3}\eta G_{Q}; \quad\quad
G_{M}=-F_{2}; \quad\quad
G_{Q}=F_{1}+F_{2}+(1+\eta)F_{3}.
\end{align}
Among various calculation prescriptions~\cite{Grach:1983hd,Brodsky:1992px,Chung:1988my,Frankfurt:1993ut}, we employ the Brodsky-Hiller (BH) framework due to its natural compatibility with our light-front approach, agreement with lattice QCD~\cite{Gurjar:2024wpq}, and proper treatment of zero-mode effects through the $I_{0,0}$ overlap.  However, if zero-mode contributions are explicitly included, all prescriptions yield equivalent results~\cite{Choi:2004ww,deMelo:1997hh}. The BH form factors are expressed as:
\begin{equation}
\begin{aligned}
G_{\mathrm{C}}^{\rm BH} &= \frac{1}{2P^{+}(1+2\eta)} \left[\frac{3-2\eta}{3}I_{0, 0}^{+} +\frac{16}{3}\eta \frac{I_{1,0}^{+}}{\sqrt{2\eta}}+\frac{2}{3}(2\eta-1)I_{1,-1}^{+}\right], \\
G_{\mathrm{M}}^{\rm BH} &= \frac{2}{2P^{+}(1+2\eta)}\left[I_{0,0}^{+}+\frac{(2\eta-1)}{\sqrt{2\eta}}I_{1,0}^{+}-I_{1,-1}^{+}\right], \\
G_{\mathrm{Q}}^{\rm BH} &= -\frac{1}{2P^{+}(1+2\eta)}\left[I_{0,0}^{+}-2\frac{I_{1,0}^{+}}{\sqrt{2\eta}}+\frac{1+\eta}{\eta}I_{1,-1}^{+} \right].  
\end{aligned}
\label{eq:ff_vector}
\end{equation}
 At $Q^2 = 0$, these form factors define fundamental observables: the electric charge $e$, magnetic moment $\mu$, and quadrupole moment $\mathcal{Q}$, as follows~\cite{Hernandez-Pinto:2024kwg,Xu:2019ilh}:
\begin{align}
eG_{\mathrm{C}}(0)=e; \quad\quad
G_{\mathrm{M}}(0)=\mu; \quad\quad
G_{\mathrm{Q}}(0)=\mathcal{Q},
\end{align}
while the electromagnetic radii of the deuteron are determined from~\cite{Chung:1988my}:
\begin{equation}
	\langle r^2_{\lbrace C,M \rbrace} \rangle= -\frac{6}{G_{\lbrace C, M\rbrace} (0)} \frac{{\rm d} G_{\lbrace C,M\rbrace}}{{\rm d}Q^2} \Big\vert_{Q^2=0}\;.
\label{eq:ff_obs}
\end{equation}
\begin{figure}
		\centering	
	\includegraphics[scale=0.35]{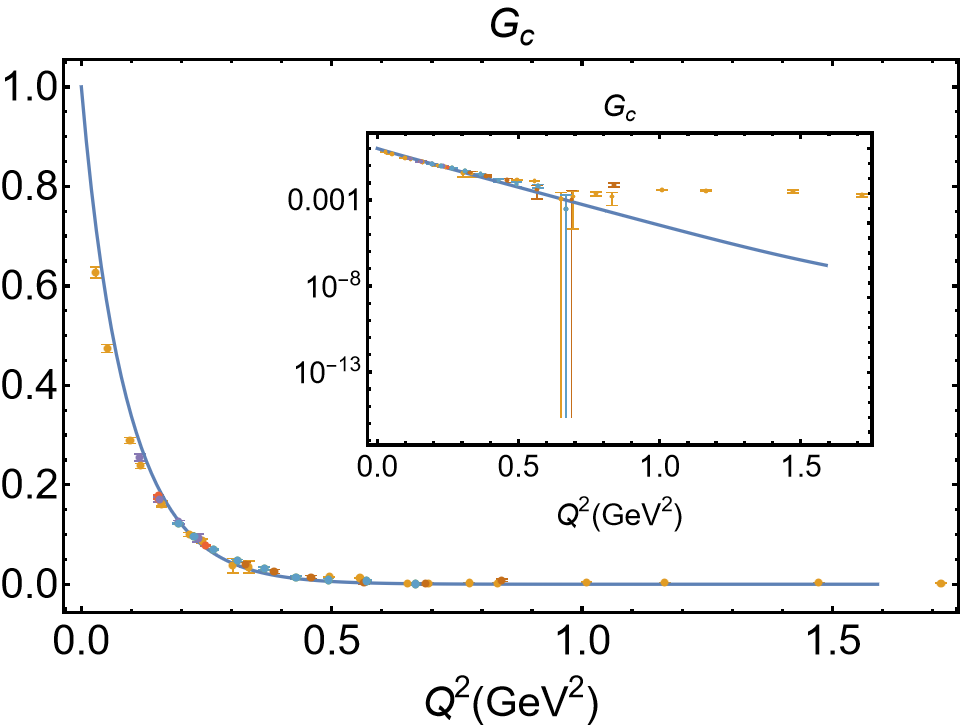}\quad\quad
	\includegraphics[scale=0.35]{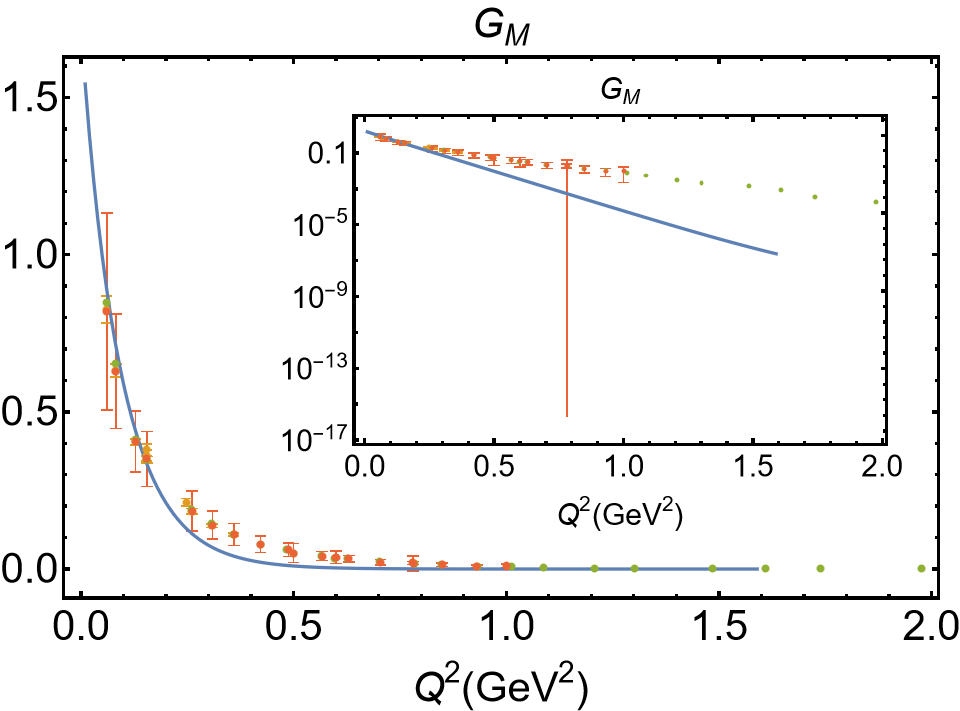}
    \includegraphics[scale=0.35]{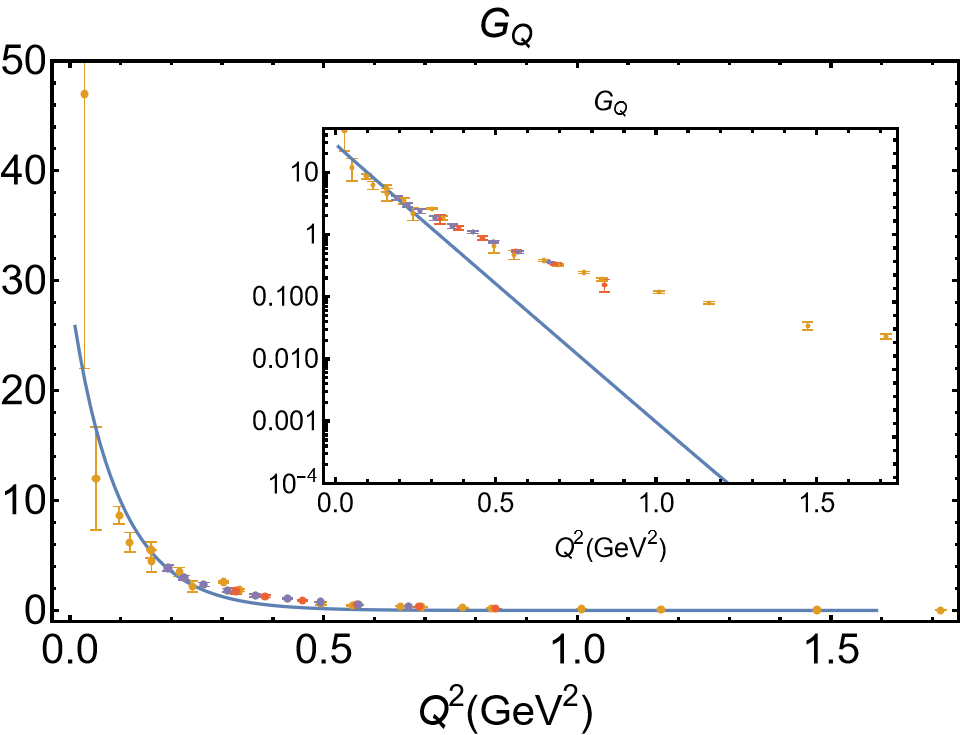}	
    \caption{The left, middle, and bottom panels show the variations of the electric, magnetic, and quadrupole form factors of the deuteron as functions of the squared momentum transfer, $Q^2$. Our results, obtained with the basis truncations $N_{\text{max}} = 8$ and $K = 9$, are compared with the available experimental data~\cite{JLABt20:2000qyq, Zhang:2011zu, The:1991eg, Benaksas:1966zz, Nikolenko:2003zq}. The insets display the form factors on a logarithmic scale.}
	\label{Fig:EMFFs}
\end{figure}

We present the variation of the charge, magnetic, and quadrupole elastic form factors with $Q^2$ in Fig.~\ref{Fig:EMFFs}, comparing our results with available experimental measurements~\cite{JLABt20:2000qyq, Zhang:2011zu, The:1991eg, Benaksas:1966zz, Nikolenko:2003zq}. Our results show good agreement with experimental data at low $Q^2$, but deviations appear as $Q^2$ increases. This discrepancy may be due to the absence of the $D$-wave and the minimal contribution of the $P$-wave in our current calculations. 

We obtain the deuteron charge radius as $\sqrt{\langle r^2_{C} \rangle} = 1.66$ fm, compared to the experimental value of $2.130 \pm 0.003 \pm 0.009$ fm~\cite{JLABt20:2000qyq}. Similarly, our approach yields a magnetic radius of $\sqrt{\langle r^2_{M} \rangle} = 1.64$ fm, compared to the experimental measurement of $1.90 \pm 0.14$ fm~\cite{Afanasev:1998hu}. Both radii are slightly underestimated in our calculations.

%We obtain the deuteron charge radius as $\sqrt{\langle r^2_{C} \rangle} = 1.66$ fm, compared to the experimental value of $2.130 \pm 0.003 \pm 0.009$ fm~\cite{JLABt20:2000qyq}. Similarly, our approach yields a magnetic radius of $\sqrt{\langle r^2_{M} \rangle} = 1.64$ fm, which is slightly underestimated relative to the experimental measuremen of $1.90 \pm 0.14$ fm~\cite{Afanasev:1998hu}.

\section{Conclusion}
We solved the LFQCD Hamiltonian for the deuteron within a combined Fock space framework, incorporating the six-quark ($|qqq\,qqq\rangle$) and six-quark–one-gluon ($|qqq\,qqq\,g\rangle$) sectors using the BLFQ approach. This allowed us to extract the deuteron's LFWFs, providing a foundation for investigating its partonic structure.

We investigated the color structure of the deuteron and analyzed the probability distribution of its various color configurations. Our results indicate that while the conventional singlet-singlet state contributes significantly, hidden-color states play a dominant role in the deuteron wave function. This finding reinforces the notion that a purely nucleon-based description is insufficient to fully capture the deuteron's internal structure. Using the extracted LFWFs, we computed the electromagnetic form factors and found good agreement with experimental data at low $Q^2$. However, deviations emerge at higher $Q^2$, which may be attributed to the absence of the $D$-wave and the minimal contribution of the $P$-wave in our current calculations.

\section*{Acknowledgement}
CM is supported by new faculty start-up funding by the Institute of Modern Physics, 
Chinese Academy of Sciences, Grant No.~E129952YR0. SK is supported by the Research Fund for International Young Scientists, 
Grant No. 12250410251, from the National Natural Science Foundation of China (NSFC), 
and the China Postdoctoral Science Foundation (CPSF), Grant No.~E339951SR0. 
XZ is supported by new faculty start-up funding by the Institute of Modern Physics, 
Chinese Academy of Sciences, by the Key Research Program of Frontier Sciences, 
Chinese Academy of Sciences, Grant No.~ZDBS-LY-7020, by the Natural Science Foundation 
of Gansu Province, China, Grant No.~20JR10RA067, by the Foundation for Key Talents 
of Gansu Province, by the Central Funds Guiding the Local Science and Technology 
Development of Gansu Province, Grant No.~22ZY1QA006, by the International Partnership 
Program of the Chinese Academy of Sciences, Grant No.~016GJHZ2022103FN, by the Strategic 
Priority Research Program of the Chinese Academy of Sciences, Grant No.~XDB34000000, 
and by the National Natural Science Foundation of China under Grant No.~12375143. 
JPV acknowledges partial support from the Department of Energy under Grant No.~DE-SC0023692. 
This research used resources of the National Energy Research Scientific Computing Center (NERSC), 
a U.S. Department of Energy Office of Science User Facility located at Lawrence Berkeley National Laboratory, 
operated under Contract No.~DE-AC02-05CH11231 using NERSC award NP-ERCAP0028672. 
A portion of the computational resources were also provided by the Gansu Computing Center. 
This research is supported by the Gansu International Collaboration and Talents Recruitment Base 
of Particle Physics, by the Senior Scientist Program funded by Gansu Province Grant No.~25RCKA008, and the International Partnership Program of the Chinese Academy of Sciences, 
Grant No.~016GJHZ2022103FN.

\biboptions{sort&compress}
\bibliographystyle{elsarticle-num}
\bibliography{ref_deuteron}      % Name of your .bib file (omit .bib extension)

\begin{thebibliography}{10}
\expandafter\ifx\csname url\endcsname\relax
  \def\url#1{\texttt{#1}}\fi
\expandafter\ifx\csname urlprefix\endcsname\relax\def\urlprefix{URL }\fi
\expandafter\ifx\csname href\endcsname\relax
  \def\href#1#2{#2} \def\path#1{#1}\fi

\bibitem{Matveev:1977xt}
V.~A. Matveev, P.~Sorba, {Is Deuteron a Six Quark System?}, Lett. Nuovo Cim. 20
  (1977) 435.
\newblock \href {https://doi.org/10.1007/BF02790723}
  {\path{doi:10.1007/BF02790723}}.

\bibitem{Matveev:1977ha}
V.~A. Matveev, P.~Sorba, {Quark Analysis of Multi - Baryonic Systems}, Nuovo
  Cim. A 45 (1978) 257.
\newblock \href {https://doi.org/10.1007/BF02724667}
  {\path{doi:10.1007/BF02724667}}.

\bibitem{Hogaasen:1979qa}
H.~Hogaasen, P.~Sorba, R.~Viollier, {Hidden Color and the Isobar Content of the
  Deuteron}, Z. Phys. C 4 (1980) 131.
\newblock \href {https://doi.org/10.1007/BF01554396}
  {\path{doi:10.1007/BF01554396}}.

\bibitem{Brodsky:1983vf}
S.~J. Brodsky, C.-R. Ji, G.~P. Lepage, {Quantum Chromodynamic Predictions for
  the Deuteron Form-Factor}, Phys. Rev. Lett. 51 (1983) 83.
\newblock \href {https://doi.org/10.1103/PhysRevLett.51.83}
  {\path{doi:10.1103/PhysRevLett.51.83}}.

\bibitem{Farrar:1991qi}
G.~R. Farrar, K.~Huleihel, H.-y. Zhang, {Deuteron form-factor}, Phys. Rev.
  Lett. 74 (1995) 650--653.
\newblock \href {https://doi.org/10.1103/PhysRevLett.74.650}
  {\path{doi:10.1103/PhysRevLett.74.650}}.

\bibitem{Bashkanov:2013cla}
M.~Bashkanov, S.~J. Brodsky, H.~Clement, {Novel Six-Quark Hidden-Color Dibaryon
  States in QCD}, Phys. Lett. B 727 (2013) 438--442.
\newblock \href {http://arxiv.org/abs/1308.6404} {\path{arXiv:1308.6404}},
  \href {https://doi.org/10.1016/j.physletb.2013.10.059}
  {\path{doi:10.1016/j.physletb.2013.10.059}}.

\bibitem{Miller:2013hla}
G.~A. Miller, {Pionic and Hidden-Color, Six-Quark Contributions to the Deuteron
  b1 Structure Function}, Phys. Rev. C 89~(4) (2014) 045203.
\newblock \href {http://arxiv.org/abs/1311.4561} {\path{arXiv:1311.4561}},
  \href {https://doi.org/10.1103/PhysRevC.89.045203}
  {\path{doi:10.1103/PhysRevC.89.045203}}.

\bibitem{Bakker:2014cua}
B.~L.~G. Bakker, C.-R. Ji, {Nuclear chromodynamics: Novel nuclear phenomena
  predicted by QCD}, Prog. Part. Nucl. Phys. 74 (2014) 1--34.
\newblock \href {https://doi.org/10.1016/j.ppnp.2013.10.001}
  {\path{doi:10.1016/j.ppnp.2013.10.001}}.

\bibitem{Gutsche:2016lrz}
T.~Gutsche, V.~E. Lyubovitskij, I.~Schmidt, {Deuteron electromagnetic structure
  functions and polarization properties in soft-wall AdS/QCD}, Phys. Rev. D
  94~(11) (2016) 116006.
\newblock \href {http://arxiv.org/abs/1607.04124} {\path{arXiv:1607.04124}},
  \href {https://doi.org/10.1103/PhysRevD.94.116006}
  {\path{doi:10.1103/PhysRevD.94.116006}}.

\bibitem{Gutsche:2015xva}
T.~Gutsche, V.~E. Lyubovitskij, I.~Schmidt, A.~Vega, {Nuclear physics in
  soft-wall AdS/QCD: Deuteron electromagnetic form factors}, Phys. Rev. D
  91~(11) (2015) 114001.
\newblock \href {http://arxiv.org/abs/1501.02738} {\path{arXiv:1501.02738}},
  \href {https://doi.org/10.1103/PhysRevD.91.114001}
  {\path{doi:10.1103/PhysRevD.91.114001}}.

\bibitem{Sun:2016ncc}
B.-d. Sun, Y.-b. Dong, {Deuteron electromagnetic form factors with the
  light-front approach}, Chin. Phys. C 41~(1) (2017) 013102.
\newblock \href {http://arxiv.org/abs/1605.02517} {\path{arXiv:1605.02517}},
  \href {https://doi.org/10.1088/1674-1137/41/1/013102}
  {\path{doi:10.1088/1674-1137/41/1/013102}}.

\bibitem{Shi:2022blm}
W.~Shi, R.~Peng, T.-X. Liu, S.~Lyu, B.~Long, {Perturbative calculations of
  deuteron form factors}, Phys. Rev. C 106~(1) (2022) 015505.
\newblock \href {http://arxiv.org/abs/2205.02000} {\path{arXiv:2205.02000}},
  \href {https://doi.org/10.1103/PhysRevC.106.015505}
  {\path{doi:10.1103/PhysRevC.106.015505}}.

\bibitem{Huseynova:2022tok}
N.~Huseynova, S.~Mamedov, J.~Samadov, {Deuteron electromagnetic form factors
  and tensor polarization observables in the framework of the hard-wall AdS/QCD
  model}, Chin. Phys. C 47 (2023) 013104.
\newblock \href {http://arxiv.org/abs/2204.06205} {\path{arXiv:2204.06205}},
  \href {https://doi.org/10.1088/1674-1137/ac957a}
  {\path{doi:10.1088/1674-1137/ac957a}}.

\bibitem{Marcucci:2015rca}
L.~E. Marcucci, F.~Gross, M.~T. Pena, M.~Piarulli, R.~Schiavilla, I.~Sick,
  A.~Stadler, J.~W. Van~Orden, M.~Viviani, {Electromagnetic Structure of
  Few-Nucleon Ground States}, J. Phys. G 43 (2016) 023002.
\newblock \href {http://arxiv.org/abs/1504.05063} {\path{arXiv:1504.05063}},
  \href {https://doi.org/10.1088/0954-3899/43/2/023002}
  {\path{doi:10.1088/0954-3899/43/2/023002}}.

\bibitem{Khachi:2023iqp}
A.~Khachi, L.~Kumar, M.~R.~G. Kumar, O.~S. K.~S. Sastri, {Deuteron structure
  and form factors: Using an inverse potential approach}, Phys. Rev. C 107~(6)
  (2023) 064002.
\newblock \href {https://doi.org/10.1103/PhysRevC.107.064002}
  {\path{doi:10.1103/PhysRevC.107.064002}}.

\bibitem{Chemtob:1974nf}
M.~Chemtob, E.~J. Moniz, M.~Rho, {Deuteron Electromagnetic Structure at Large
  Momentum Transfer}, Phys. Rev. C 10 (1974) 344--352.
\newblock \href {https://doi.org/10.1103/PhysRevC.10.344}
  {\path{doi:10.1103/PhysRevC.10.344}}.

\bibitem{Blankenbecler:1971xa}
R.~Blankenbecler, J.~F. Gunion, {THE DEUTERON ELECTROMAGNETIC FORM-FACTOR},
  Phys. Rev. D 4 (1971) 718.
\newblock \href {https://doi.org/10.1103/PhysRevD.4.718}
  {\path{doi:10.1103/PhysRevD.4.718}}.

\bibitem{Piller:1995mf}
G.~Piller, W.~Melnitchouk, A.~W. Thomas, {Polarized deep inelastic scattering
  from nuclei: A Relativistic approach}, Phys. Rev. C 54 (1996) 894--903.
\newblock \href {http://arxiv.org/abs/nucl-th/9605045}
  {\path{arXiv:nucl-th/9605045}}, \href
  {https://doi.org/10.1103/PhysRevC.54.894}
  {\path{doi:10.1103/PhysRevC.54.894}}.

\bibitem{JLABt20:2000qyq}
D.~Abbott, et~al., {Phenomenology of the deuteron electromagnetic
  form-factors}, Eur. Phys. J. A 7 (2000) 421--427.
\newblock \href {http://arxiv.org/abs/nucl-ex/0002003}
  {\path{arXiv:nucl-ex/0002003}}, \href {https://doi.org/10.1007/PL00013629}
  {\path{doi:10.1007/PL00013629}}.

\bibitem{Sick:1998cvq}
I.~Sick, D.~Trautmann, {On the rms radius of the deuteron}, Nucl. Phys. A 637
  (1998) 559--575.
\newblock \href {https://doi.org/10.1016/S0375-9474(98)00334-0}
  {\path{doi:10.1016/S0375-9474(98)00334-0}}.

\bibitem{Zhang:2011zu}
C.~Zhang, et~al., {Precise Measurement of Deuteron Tensor Analyzing Powers with
  BLAST}, Phys. Rev. Lett. 107 (2011) 252501.
\newblock \href {https://doi.org/10.1103/PhysRevLett.107.252501}
  {\path{doi:10.1103/PhysRevLett.107.252501}}.

\bibitem{Nikolenko:2003zq}
D.~M. Nikolenko, et~al., {Measurement of the tensor analyzing powers T(20) and
  T(21) in elastic electron deuteron scattering}, Phys. Rev. Lett. 90 (2003)
  072501.
\newblock \href {https://doi.org/10.1103/PhysRevLett.90.072501}
  {\path{doi:10.1103/PhysRevLett.90.072501}}.

\bibitem{Garcon:2001sz}
M.~Garcon, J.~W. Van~Orden, {The Deuteron: Structure and form-factors}, Adv.
  Nucl. Phys. 26 (2001) 293.
\newblock \href {http://arxiv.org/abs/nucl-th/0102049}
  {\path{arXiv:nucl-th/0102049}}, \href
  {https://doi.org/10.1007/0-306-47915-X_4}
  {\path{doi:10.1007/0-306-47915-X_4}}.

\bibitem{Chen:2024rgi}
C.~Chen, L.~Liu, P.~Sun, Y.-B. Yang, Y.~Geng, F.~Yao, J.-H. Zhang, K.~Zhang,
  {Parton Distribution Function of a Deuteron-like Dibaryon System from Lattice
  QCD} (8 2024).
\newblock \href {http://arxiv.org/abs/2408.12819} {\path{arXiv:2408.12819}}.

\bibitem{HERMES:2005pon}
A.~Airapetian, et~al., {First measurement of the tensor structure function b(1)
  of the deuteron}, Phys. Rev. Lett. 95 (2005) 242001.
\newblock \href {http://arxiv.org/abs/hep-ex/0506018}
  {\path{arXiv:hep-ex/0506018}}, \href
  {https://doi.org/10.1103/PhysRevLett.95.242001}
  {\path{doi:10.1103/PhysRevLett.95.242001}}.

\bibitem{ProposalJLab2023}
K.~Slifer, et~al., {The Deuteron Tensor Structure Function b1 (Report No.
  E12-13-011)},
  \url{https://www.jlab.org/sites/default/files/PAC/PAC51/PAC51_REPORT_2023.pdf}.

\bibitem{AbilityFermiLabE1039}
M.~Brooks, et~al., {SeaQuest with a Transversely Polarized Target (E1039)},
  \url{https://twist.phys.virginia.edu/work/E1039proposal_final.pdf}.

\bibitem{Keller:2022abm}
D.~Keller, {The Transverse Structure of the Deuteron with Drell-Yan} (5 2022).
\newblock \href {http://arxiv.org/abs/2205.01249} {\path{arXiv:2205.01249}}.

\bibitem{Hoodbhoy:1988am}
P.~Hoodbhoy, R.~L. Jaffe, A.~Manohar, {Novel Effects in Deep Inelastic
  Scattering from Spin 1 Hadrons}, Nucl. Phys. B 312 (1989) 571--588.
\newblock \href {https://doi.org/10.1016/0550-3213(89)90572-5}
  {\path{doi:10.1016/0550-3213(89)90572-5}}.

\bibitem{Khan:1991qk}
H.~Khan, P.~Hoodbhoy, {Convenient parametrization for deep inelastic structure
  functions of the deuteron}, Phys. Rev. C 44 (1991) 1219--1222.
\newblock \href {https://doi.org/10.1103/PhysRevC.44.1219}
  {\path{doi:10.1103/PhysRevC.44.1219}}.

\bibitem{Edelmann:1997ik}
J.~Edelmann, G.~Piller, W.~Weise, {Deuteron spin structure functions at small
  Bjorken x}, Phys. Rev. C 57 (1998) 3392--3405.
\newblock \href {http://arxiv.org/abs/hep-ph/9709455}
  {\path{arXiv:hep-ph/9709455}}, \href
  {https://doi.org/10.1103/PhysRevC.57.3392}
  {\path{doi:10.1103/PhysRevC.57.3392}}.

\bibitem{Cosyn:2017fbo}
W.~Cosyn, Y.-B. Dong, S.~Kumano, M.~Sargsian, {Tensor-polarized structure
  function $b_1$ in standard convolution description of deuteron}, Phys. Rev. D
  95~(7) (2017) 074036.
\newblock \href {http://arxiv.org/abs/1702.05337} {\path{arXiv:1702.05337}},
  \href {https://doi.org/10.1103/PhysRevD.95.074036}
  {\path{doi:10.1103/PhysRevD.95.074036}}.

\bibitem{Kumano:2024fpr}
S.~Kumano, {Parton distribution functions and fragmentation functions of spin-1
  hadrons} (6 2024).
\newblock \href {http://arxiv.org/abs/2406.01180} {\path{arXiv:2406.01180}}.

\bibitem{Accardi:2012qut}
A.~Accardi, et~al., {Electron Ion Collider: The Next QCD Frontier}:
  {Understanding the glue that binds us all}, Eur. Phys. J. A 52~(9) (2016)
  268.
\newblock \href {http://arxiv.org/abs/1212.1701} {\path{arXiv:1212.1701}},
  \href {https://doi.org/10.1140/epja/i2016-16268-9}
  {\path{doi:10.1140/epja/i2016-16268-9}}.

\bibitem{Anderle:2021wcy}
D.~P. Anderle, et~al., {Electron-ion collider in China}, Front. Phys. (Beijing)
  16~(6) (2021) 64701.
\newblock \href {http://arxiv.org/abs/2102.09222} {\path{arXiv:2102.09222}},
  \href {https://doi.org/10.1007/s11467-021-1062-0}
  {\path{doi:10.1007/s11467-021-1062-0}}.

\bibitem{Vary:2009gt}
J.~P. Vary, H.~Honkanen, J.~Li, P.~Maris, S.~J. Brodsky, A.~Harindranath, G.~F.
  de~Teramond, P.~Sternberg, E.~G. Ng, C.~Yang, {Hamiltonian light-front field
  theory in a basis function approach}, Phys. Rev. C 81 (2010) 035205.
\newblock \href {http://arxiv.org/abs/0905.1411} {\path{arXiv:0905.1411}},
  \href {https://doi.org/10.1103/PhysRevC.81.035205}
  {\path{doi:10.1103/PhysRevC.81.035205}}.

\bibitem{Lan:2019vui}
J.~Lan, C.~Mondal, S.~Jia, X.~Zhao, J.~P. Vary, {Parton Distribution Functions
  from a Light Front Hamiltonian and QCD Evolution for Light Mesons}, Phys.
  Rev. Lett. 122~(17) (2019) 172001.
\newblock \href {http://arxiv.org/abs/1901.11430} {\path{arXiv:1901.11430}},
  \href {https://doi.org/10.1103/PhysRevLett.122.172001}
  {\path{doi:10.1103/PhysRevLett.122.172001}}.

\bibitem{Lan:2019rba}
J.~Lan, C.~Mondal, S.~Jia, X.~Zhao, J.~P. Vary, {Pion and kaon parton
  distribution functions from basis light front quantization and QCD
  evolution}, Phys. Rev. D 101~(3) (2020) 034024.
\newblock \href {http://arxiv.org/abs/1907.01509} {\path{arXiv:1907.01509}},
  \href {https://doi.org/10.1103/PhysRevD.101.034024}
  {\path{doi:10.1103/PhysRevD.101.034024}}.

\bibitem{Mondal:2019jdg}
C.~Mondal, S.~Xu, J.~Lan, X.~Zhao, Y.~Li, D.~Chakrabarti, J.~P. Vary, {Proton
  structure from a light-front Hamiltonian}, Phys. Rev. D 102~(1) (2020)
  016008.
\newblock \href {http://arxiv.org/abs/1911.10913} {\path{arXiv:1911.10913}},
  \href {https://doi.org/10.1103/PhysRevD.102.016008}
  {\path{doi:10.1103/PhysRevD.102.016008}}.

\bibitem{Lan:2019img}
J.~Lan, C.~Mondal, M.~Li, Y.~Li, S.~Tang, X.~Zhao, J.~P. Vary, {Parton
  Distribution Functions of Heavy Mesons on the Light Front}, Phys. Rev. D
  102~(1) (2020) 014020.
\newblock \href {http://arxiv.org/abs/1911.11676} {\path{arXiv:1911.11676}},
  \href {https://doi.org/10.1103/PhysRevD.102.014020}
  {\path{doi:10.1103/PhysRevD.102.014020}}.

\bibitem{Lan:2021wok}
J.~Lan, K.~Fu, C.~Mondal, X.~Zhao, j.~P. Vary, {Light mesons with one dynamical
  gluon on the light front}, Phys. Lett. B 825 (2022) 136890.
\newblock \href {http://arxiv.org/abs/2106.04954} {\path{arXiv:2106.04954}},
  \href {https://doi.org/10.1016/j.physletb.2022.136890}
  {\path{doi:10.1016/j.physletb.2022.136890}}.

\bibitem{Xu:2021wwj}
S.~Xu, C.~Mondal, J.~Lan, X.~Zhao, Y.~Li, J.~P. Vary, {Nucleon structure from
  basis light-front quantization}, Phys. Rev. D 104~(9) (2021) 094036.
\newblock \href {http://arxiv.org/abs/2108.03909} {\path{arXiv:2108.03909}},
  \href {https://doi.org/10.1103/PhysRevD.104.094036}
  {\path{doi:10.1103/PhysRevD.104.094036}}.

\bibitem{Liu:2022fvl}
Y.~Liu, S.~Xu, C.~Mondal, X.~Zhao, J.~P. Vary, {Angular momentum and
  generalized parton distributions for the proton with basis light-front
  quantization}, Phys. Rev. D 105~(9) (2022) 094018.
\newblock \href {http://arxiv.org/abs/2202.00985} {\path{arXiv:2202.00985}},
  \href {https://doi.org/10.1103/PhysRevD.105.094018}
  {\path{doi:10.1103/PhysRevD.105.094018}}.

\bibitem{Hu:2022ctr}
Z.~Hu, S.~Xu, C.~Mondal, X.~Zhao, J.~P. Vary, {Transverse momentum structure of
  proton within the basis light-front quantization framework}, Phys. Lett. B
  833 (2022) 137360.
\newblock \href {http://arxiv.org/abs/2205.04714} {\path{arXiv:2205.04714}},
  \href {https://doi.org/10.1016/j.physletb.2022.137360}
  {\path{doi:10.1016/j.physletb.2022.137360}}.

\bibitem{Peng:2022lte}
T.~Peng, Z.~Zhu, S.~Xu, X.~Liu, C.~Mondal, X.~Zhao, J.~P. Vary, {Basis
  light-front quantization approach to \ensuremath{\Lambda} and
  \ensuremath{\Lambda}c and their isospin triplet baryons}, Phys. Rev. D
  106~(11) (2022) 114040.
\newblock \href {http://arxiv.org/abs/2208.00355} {\path{arXiv:2208.00355}},
  \href {https://doi.org/10.1103/PhysRevD.106.114040}
  {\path{doi:10.1103/PhysRevD.106.114040}}.

\bibitem{Xu:2022yxb}
S.~Xu, C.~Mondal, X.~Zhao, Y.~Li, J.~P. Vary, {Quark and gluon spin and orbital
  angular momentum in the proton}, Phys. Rev. D 108~(9) (2023) 094002.
\newblock \href {http://arxiv.org/abs/2209.08584} {\path{arXiv:2209.08584}},
  \href {https://doi.org/10.1103/PhysRevD.108.094002}
  {\path{doi:10.1103/PhysRevD.108.094002}}.

\bibitem{Zhu:2023lst}
Z.~Zhu, Z.~Hu, J.~Lan, C.~Mondal, X.~Zhao, J.~P. Vary, {Transverse structure of
  the pion beyond leading twist with basis light-front quantization}, Phys.
  Lett. B 839 (2023) 137808.
\newblock \href {http://arxiv.org/abs/2301.12994} {\path{arXiv:2301.12994}},
  \href {https://doi.org/10.1016/j.physletb.2023.137808}
  {\path{doi:10.1016/j.physletb.2023.137808}}.

\bibitem{Zhu:2023nhl}
Z.~Zhu, T.~Peng, Z.~Hu, S.~Xu, C.~Mondal, X.~Zhao, J.~P. Vary, {Transverse
  momentum structure of strange and charmed baryons: A light-front Hamiltonian
  approach}, Phys. Rev. D 108~(3) (2023) 036009.
\newblock \href {http://arxiv.org/abs/2304.05058} {\path{arXiv:2304.05058}},
  \href {https://doi.org/10.1103/PhysRevD.108.036009}
  {\path{doi:10.1103/PhysRevD.108.036009}}.

\bibitem{Kaur:2023lun}
S.~Kaur, S.~Xu, C.~Mondal, X.~Zhao, J.~P. Vary, {Spatial imaging of proton via
  leading-twist nonskewed GPDs with basis light-front quantization}, Phys. Rev.
  D 109~(1) (2024) 014015.
\newblock \href {http://arxiv.org/abs/2307.09869} {\path{arXiv:2307.09869}},
  \href {https://doi.org/10.1103/PhysRevD.109.014015}
  {\path{doi:10.1103/PhysRevD.109.014015}}.

\bibitem{Lin:2023ezw}
B.~Lin, S.~Nair, S.~Xu, Z.~Hu, C.~Mondal, X.~Zhao, J.~P. Vary, {Generalized
  parton distributions of gluon in proton: A light-front quantization
  approach}, Phys. Lett. B 847 (2023) 138305.
\newblock \href {http://arxiv.org/abs/2308.08275} {\path{arXiv:2308.08275}},
  \href {https://doi.org/10.1016/j.physletb.2023.138305}
  {\path{doi:10.1016/j.physletb.2023.138305}}.

\bibitem{Zhang:2023xfe}
Z.~Zhang, Z.~Hu, S.~Xu, C.~Mondal, X.~Zhao, J.~P. Vary, {Twist-3 generalized
  parton distribution for the proton from basis light-front quantization},
  Phys. Rev. D 109~(3) (2024) 034031.
\newblock \href {http://arxiv.org/abs/2312.00667} {\path{arXiv:2312.00667}},
  \href {https://doi.org/10.1103/PhysRevD.109.034031}
  {\path{doi:10.1103/PhysRevD.109.034031}}.

\bibitem{Kaur:2024iwn}
S.~Kaur, J.~Wu, Z.~Hu, J.~Lan, C.~Mondal, X.~Zhao, J.~P. Vary, {Quark and gluon
  distributions in \ensuremath{\rho}-meson from basis light-front
  quantization}, Phys. Lett. B 851 (2024) 138563.
\newblock \href {http://arxiv.org/abs/2401.03480} {\path{arXiv:2401.03480}},
  \href {https://doi.org/10.1016/j.physletb.2024.138563}
  {\path{doi:10.1016/j.physletb.2024.138563}}.

\bibitem{Liu:2024umn}
Y.~Liu, S.~Xu, C.~Mondal, Z.~Hu, X.~Zhao, J.~P. Vary, {Skewed generalized
  parton distributions of proton from basis light-front quantization}, Phys.
  Lett. B 855 (2024) 138809.
\newblock \href {http://arxiv.org/abs/2403.05922} {\path{arXiv:2403.05922}},
  \href {https://doi.org/10.1016/j.physletb.2024.138809}
  {\path{doi:10.1016/j.physletb.2024.138809}}.

\bibitem{Yu:2024mxo}
H.~Yu, Z.~Hu, S.~Xu, C.~Mondal, X.~Zhao, J.~P. Vary,
  {Transverse-momentum-dependent gluon distributions of proton within basis
  light-front quantization}, Phys. Lett. B 855 (2024) 138831.
\newblock \href {http://arxiv.org/abs/2403.06125} {\path{arXiv:2403.06125}},
  \href {https://doi.org/10.1016/j.physletb.2024.138831}
  {\path{doi:10.1016/j.physletb.2024.138831}}.

\bibitem{Nair:2024fit}
S.~Nair, C.~Mondal, S.~Xu, X.~Zhao, A.~Mukherjee, J.~P. Vary, {Gravitational
  form factors and mechanical properties of quarks in protons: A basis
  light-front quantization approach}, Phys. Rev. D 110~(5) (2024) 056027.
\newblock \href {http://arxiv.org/abs/2403.11702} {\path{arXiv:2403.11702}},
  \href {https://doi.org/10.1103/PhysRevD.110.056027}
  {\path{doi:10.1103/PhysRevD.110.056027}}.

\bibitem{Zhu:2024awq}
Z.~Zhu, S.~Xu, J.~Wu, H.~Yu, Z.~Hu, J.~Lan, C.~Mondal, X.~Zhao, J.~P. Vary,
  {Transverse structure of the proton beyond leading twist: A light-front
  Hamiltonian approach}, Phys. Lett. B 855 (2024) 138829.
\newblock \href {http://arxiv.org/abs/2404.13720} {\path{arXiv:2404.13720}},
  \href {https://doi.org/10.1016/j.physletb.2024.138829}
  {\path{doi:10.1016/j.physletb.2024.138829}}.

\bibitem{Wu:2024hre}
X.~Wu, Z.~Zhu, Z.~Lin, C.~Mondal, J.~Lan, X.~Zhao, J.~P. Vary, {Pion to photon
  transition form factor: Beyond valence quarks} (8 2024).
\newblock \href {http://arxiv.org/abs/2408.06871} {\path{arXiv:2408.06871}}.

\bibitem{Lin:2024ijo}
B.~Lin, S.~Nair, C.~Mondal, S.~Xu, Z.~Hu, P.~Zhang, X.~Zhao, J.~P. Vary,
  {Chiral-odd gluon generalized parton distributions in the proton: A
  light-front quantization approach}, Phys. Lett. B 860 (2025) 139153.
\newblock \href {http://arxiv.org/abs/2408.09988} {\path{arXiv:2408.09988}},
  \href {https://doi.org/10.1016/j.physletb.2024.139153}
  {\path{doi:10.1016/j.physletb.2024.139153}}.

\bibitem{Xu:2024sjt}
S.~Xu, Y.~Liu, C.~Mondal, J.~Lan, X.~Zhao, Y.~Li, J.~P. Vary, {Towards a first
  principles light-front Hamiltonian for the nucleon} (8 2024).
\newblock \href {http://arxiv.org/abs/2408.11298} {\path{arXiv:2408.11298}}.

\bibitem{Peng:2024qpw}
T.-C. Peng, Z.~Hu, S.~Nair, S.~Xu, X.~Liu, C.~Mondal, X.~Zhao, J.~P. Vary,
  {Double parton distributions of the proton from basis light-front
  quantization} (10 2024).
\newblock \href {http://arxiv.org/abs/2410.11574} {\path{arXiv:2410.11574}}.

\bibitem{Lan:2025fia}
J.~Lan, J.~Chen, Z.~Zhu, C.~Mondal, X.~Zhao, J.~P. Vary, {Strange mesons with
  one dynamical gluon: A light-front approach} (1 2025).
\newblock \href {http://arxiv.org/abs/2501.03476} {\path{arXiv:2501.03476}}.

\bibitem{Zhang:2025nll}
P.~Zhang, Y.~Liu, S.~Xu, C.~Mondal, X.~Zhao, J.~P. Vary, {Gluon skewed
  generalized parton distributions of proton from a light-front Hamiltonian
  approach} (1 2025).
\newblock \href {http://arxiv.org/abs/2501.10119} {\path{arXiv:2501.10119}}.

\bibitem{Brodsky:1997de}
S.~J. Brodsky, H.-C. Pauli, S.~S. Pinsky, {Quantum chromodynamics and other
  field theories on the light cone}, Phys. Rept. 301 (1998) 299--486.
\newblock \href {http://arxiv.org/abs/hep-ph/9705477}
  {\path{arXiv:hep-ph/9705477}}, \href
  {https://doi.org/10.1016/S0370-1573(97)00089-6}
  {\path{doi:10.1016/S0370-1573(97)00089-6}}.

\bibitem{Glazek:1993rc}
S.~D. Glazek, K.~G. Wilson, {Renormalization of Hamiltonians}, Phys. Rev. D 48
  (1993) 5863--5872.
\newblock \href {https://doi.org/10.1103/PhysRevD.48.5863}
  {\path{doi:10.1103/PhysRevD.48.5863}}.

\bibitem{Karmanov:2008br}
V.~A. Karmanov, J.~F. Mathiot, A.~V. Smirnov, {Systematic renormalization
  scheme in light-front dynamics with Fock space truncation}, Phys. Rev. D 77
  (2008) 085028.
\newblock \href {http://arxiv.org/abs/0801.4507} {\path{arXiv:0801.4507}},
  \href {https://doi.org/10.1103/PhysRevD.77.085028}
  {\path{doi:10.1103/PhysRevD.77.085028}}.

\bibitem{Karmanov:2012aj}
V.~A. Karmanov, J.~F. Mathiot, A.~V. Smirnov, {Ab initio nonperturbative
  calculation of physical observables in light-front dynamics. Application to
  the Yukawa model}, Phys. Rev. D 86 (2012) 085006.
\newblock \href {http://arxiv.org/abs/1204.3257} {\path{arXiv:1204.3257}},
  \href {https://doi.org/10.1103/PhysRevD.86.085006}
  {\path{doi:10.1103/PhysRevD.86.085006}}.

\bibitem{Zhao:2014hpa}
X.~Zhao, {Advances in Basis Light-front Quantization}, Few Body Syst. 56~(6-9)
  (2015) 257--265.
\newblock \href {http://arxiv.org/abs/1411.7748} {\path{arXiv:1411.7748}},
  \href {https://doi.org/10.1007/s00601-015-1003-y}
  {\path{doi:10.1007/s00601-015-1003-y}}.

\bibitem{Zhao:2020kuf}
X.~Zhao, K.~Fu, H.~Zhao, J.~P. Vary, {Positronium: an illustration of
  nonperturbative renormalization in a basis light-front approach}, PoS LC2019
  (2020) 090.
\newblock \href {http://arxiv.org/abs/2103.06719} {\path{arXiv:2103.06719}},
  \href {https://doi.org/10.22323/1.374.0090} {\path{doi:10.22323/1.374.0090}}.

\bibitem{Glazek:1992aq}
S.~D. Glazek, R.~J. Perry, {Special example of relativistic Hamiltonian field
  theory}, Phys. Rev. D 45 (1992) 3740--3754.
\newblock \href {https://doi.org/10.1103/PhysRevD.45.3740}
  {\path{doi:10.1103/PhysRevD.45.3740}}.

\bibitem{Burkardt:1998dd}
M.~Burkardt, {Dynamical vertex mass generation and chiral symmetry breaking on
  the light front}, Phys. Rev. D 58 (1998) 096015.
\newblock \href {http://arxiv.org/abs/hep-th/9805088}
  {\path{arXiv:hep-th/9805088}}, \href
  {https://doi.org/10.1103/PhysRevD.58.096015}
  {\path{doi:10.1103/PhysRevD.58.096015}}.

\bibitem{Arnold:1979cg}
R.~G. Arnold, C.~E. Carlson, F.~Gross, {Elastic electron-Deuteron Scattering at
  High-Energy}, Phys. Rev. C 21 (1980) 1426.
\newblock \href {https://doi.org/10.1103/PhysRevC.21.1426}
  {\path{doi:10.1103/PhysRevC.21.1426}}.

\bibitem{Cardarelli:1994yq}
F.~Cardarelli, I.~L. Grach, I.~M. Narodetsky, G.~Salme, S.~Simula,
  {Electromagnetic form-factors of the rho meson in a light front constituent
  quark model}, Phys. Lett. B 349 (1995) 393--399.
\newblock \href {http://arxiv.org/abs/hep-ph/9502360}
  {\path{arXiv:hep-ph/9502360}}, \href
  {https://doi.org/10.1016/0370-2693(95)00230-I}
  {\path{doi:10.1016/0370-2693(95)00230-I}}.

\bibitem{Brodsky:1992px}
S.~J. Brodsky, J.~R. Hiller, {Universal properties of the electromagnetic
  interactions of spin one systems}, Phys. Rev. D 46 (1992) 2141--2149.
\newblock \href {https://doi.org/10.1103/PhysRevD.46.2141}
  {\path{doi:10.1103/PhysRevD.46.2141}}.

\bibitem{Qian:2020utg}
W.~Qian, S.~Jia, Y.~Li, J.~P. Vary, {Light mesons within the basis light-front
  quantization framework}, Phys. Rev. C 102~(5) (2020) 055207.
\newblock \href {http://arxiv.org/abs/2005.13806} {\path{arXiv:2005.13806}},
  \href {https://doi.org/10.1103/PhysRevC.102.055207}
  {\path{doi:10.1103/PhysRevC.102.055207}}.

\bibitem{Li:2021cwv}
M.~Li, Y.~Li, G.~Chen, T.~Lappi, J.~P. Vary, {Light-front wavefunctions of
  mesons by design}, Eur. Phys. J. C 82~(11) (2022) 1045.
\newblock \href {http://arxiv.org/abs/2111.07087} {\path{arXiv:2111.07087}},
  \href {https://doi.org/10.1140/epjc/s10052-022-10988-5}
  {\path{doi:10.1140/epjc/s10052-022-10988-5}}.

\bibitem{Choi:2004ww}
H.-M. Choi, C.-R. Ji, {Electromagnetic structure of the rho meson in the light
  front quark model}, Phys. Rev. D 70 (2004) 053015.
\newblock \href {http://arxiv.org/abs/hep-ph/0402114}
  {\path{arXiv:hep-ph/0402114}}, \href
  {https://doi.org/10.1103/PhysRevD.70.053015}
  {\path{doi:10.1103/PhysRevD.70.053015}}.

\bibitem{Grach:1983hd}
I.~L. Grach, L.~A. Kondratyuk, {ELECTROMAGNETIC FORM-FACTOR OF DEUTERON IN
  RELATIVISTIC DYNAMICS. TWO NUCLEON AND SIX QUARK COMPONENTS}, Sov. J. Nucl.
  Phys. 39 (1984) 198.

\bibitem{Chung:1988my}
P.~L. Chung, W.~N. Polyzou, F.~Coester, B.~D. Keister, {Hamiltonian Light Front
  Dynamics of Elastic electron Deuteron Scattering}, Phys. Rev. C 37 (1988)
  2000--2015.
\newblock \href {https://doi.org/10.1103/PhysRevC.37.2000}
  {\path{doi:10.1103/PhysRevC.37.2000}}.

\bibitem{Frankfurt:1993ut}
L.~L. Frankfurt, M.~Strikman, T.~Frederico, {Deuteron form-factors in the light
  cone quantum mechanics 'good' component approach}, Phys. Rev. C 48 (1993)
  2182--2189.
\newblock \href {https://doi.org/10.1103/PhysRevC.48.2182}
  {\path{doi:10.1103/PhysRevC.48.2182}}.

\bibitem{Gurjar:2024wpq}
B.~Gurjar, C.~Mondal, S.~Kaur, {\ensuremath{\rho}-meson spectroscopy and
  diffractive production using the holographic light-front Schr\"odinger
  equation and the \textquoteright{}t Hooft equation}, Phys. Rev. D 109~(9)
  (2024) 094017.
\newblock \href {http://arxiv.org/abs/2401.13514} {\path{arXiv:2401.13514}},
  \href {https://doi.org/10.1103/PhysRevD.109.094017}
  {\path{doi:10.1103/PhysRevD.109.094017}}.

\bibitem{deMelo:1997hh}
J.~P. B.~C. de~Melo, T.~Frederico, {Covariant and light front approaches to the
  rho meson electromagnetic form-factors}, Phys. Rev. C 55 (1997) 2043.
\newblock \href {http://arxiv.org/abs/nucl-th/9706032}
  {\path{arXiv:nucl-th/9706032}}, \href
  {https://doi.org/10.1103/PhysRevC.55.2043}
  {\path{doi:10.1103/PhysRevC.55.2043}}.

\bibitem{Hernandez-Pinto:2024kwg}
R.~J. Hern\'andez-Pinto, L.~X. Guti\'errez-Guerrero, M.~A. Bedolla, A.~Bashir,
  {Electric, magnetic, and quadrupole form factors and charge radii of vector
  mesons: From light to heavy sectors in a contact interaction}, Phys. Rev. D
  110~(11) (2024) 114015.
\newblock \href {http://arxiv.org/abs/2410.23813} {\path{arXiv:2410.23813}},
  \href {https://doi.org/10.1103/PhysRevD.110.114015}
  {\path{doi:10.1103/PhysRevD.110.114015}}.

\bibitem{Xu:2019ilh}
Y.-Z. Xu, D.~Binosi, Z.-F. Cui, B.-L. Li, C.~D. Roberts, S.-S. Xu, H.~S. Zong,
  {Elastic electromagnetic form factors of vector mesons}, Phys. Rev. D
  100~(11) (2019) 114038.
\newblock \href {http://arxiv.org/abs/1911.05199} {\path{arXiv:1911.05199}},
  \href {https://doi.org/10.1103/PhysRevD.100.114038}
  {\path{doi:10.1103/PhysRevD.100.114038}}.

\bibitem{The:1991eg}
I.~The, et~al., {Measurement of tensor polarization in elastic electron
  deuteron scattering in the momentum transfer range 3.8-fm**(-1)
  \ensuremath{<}= q \ensuremath{<}= 4.6-fm**(-1)}, Phys. Rev. Lett. 67 (1991)
  173--176.
\newblock \href {https://doi.org/10.1103/PhysRevLett.67.173}
  {\path{doi:10.1103/PhysRevLett.67.173}}.

\bibitem{Benaksas:1966zz}
D.~Benaksas, D.~Drickey, D.~Frerejacque, {Deuteron Electromagnetic Form Factors
  for F-3-2 \ensuremath{<} q2 \ensuremath{<} F-6-2}, Phys. Rev. 148 (1966)
  1327--1331.
\newblock \href {https://doi.org/10.1103/PhysRev.148.1327}
  {\path{doi:10.1103/PhysRev.148.1327}}.

\bibitem{Afanasev:1998hu}
A.~Afanasev, V.~D. Afanasev, S.~V. Trubnikov, {Magnetic radius of the deuteron}
  (8 1998).
\newblock \href {http://arxiv.org/abs/nucl-th/9808047}
  {\path{arXiv:nucl-th/9808047}}.

\end{thebibliography}

 \end{document}